\documentclass[onecolumn,aps,prd,showpacs,floatfix,eqsecnum,nofootinbib]
{revtex4}
\usepackage{amsmath,epsfig}
\begin{document}
\title{Towards a formalism for mapping the spacetimes of massive
compact objects:\\ Bumpy black holes and their orbits}
\author{Nathan A.\ Collins and Scott A.\ Hughes}
\affiliation{Department of Physics, MIT, 77 Massachusetts Ave.,
Cambridge, MA 02139}
\begin{abstract}
Astronomical observations have established that extremely compact,
massive objects are common in the universe.  It is generally accepted
that these objects are, in all likelihood, black holes.  As
observational technology has improved, it has become possible to test
this hypothesis in ever greater detail.  In particular, it is or will
be possible to measure the properties of orbits deep in the strong
field of a black hole candidate (using x-ray timing or future
gravitational-wave measurements) and to test whether they have the
characteristics of black hole orbits in general relativity.  Past work
has shown that, in principle, such measurements can be used to map the
spacetime of a massive compact object, testing in particular whether
the object's multipolar structure satisfies the rather strict
constraints imposed by the black hole hypothesis.  Performing such a
test in practice requires that we be able to compare against objects
with the ``wrong'' multipole structure.  In this paper, we present
tools for constructing the spacetimes of {\it bumpy black holes}:
objects that are {\it almost}\ black holes, but that have some
multipoles with the wrong value.  In this first analysis, we focus on
objects with no angular momentum.  Generalization to bumpy Kerr black
holes should be straightforward, albeit labor intensive.  Our
construction has two particularly desirable properties.  First, the
spacetimes which we present are good deep into the strong field of the
object --- we do not use a ``large $r$'' expansion (except to make
contact with weak field intuition).  Second, our spacetimes reduce to
the exact black hole spacetimes of general relativity in a natural
way, by dialing the ``bumpiness'' of the black hole to zero.  We
propose that bumpy black holes can be used as the foundation for a
null experiment: if black hole candidates are indeed the black holes
of general relativity, their bumpiness should be zero.  By comparing
the properties of orbits in a bumpy spacetime with those measured by
an astrophysical source, observations should be able to test this
hypothesis, stringently testing whether they are in fact the black
holes of general relativity.
\end{abstract}
\maketitle

\section{Introduction}

\subsection{Motivation}
\label{subsec:motive}

Observations have now established that the cores of nearly all nearby
galaxies contain a massive, compact, dark object {\cite{kr95,k03}}.
These objects range in mass from several $10^5\,M_\odot$ to several
$10^9\,M_\odot$.  Extremely compact stellar mass objects ($\sim
10\,M_\odot$ or so) exist and have been studied in the galactic field
(see, e.g., Ref.\ {\cite{bailynetal98}} for a review).  Evidence
suggests the existence of objects with intermediate masses,
$10^2\,M_\odot - 10^4\,M_\odot$, filling the gap between the
supermassive and stellar mass objects (see, e.g., Ref.\ {\cite{mc03}}
for a review).  The most generally accepted explanation is that these
compact bodies are massive black holes.

Although this is the most generally accepted explanation for these
objects, it is not the only explanation.  In some cases, the massive
dark objects seen in galaxy cores can be explained quite well as dense
clusters of stars or stellar remnants.  Such models are rapidly
becoming disfavored in many cases as our ability to study the central
regions of galaxies improves --- many of these putative clusters would
have to be so compact that they would not be gravitationally stable.
By using ``exotic'' matter, it becomes possible to build objects that
are massive, compact, but stable.  For examine, by tuning the mass and
self interaction of a massive scalar field {\cite{csw86,r_boson}}, one
can build an object that is consistent with much of the observational
evidence available today.  Indeed, the fields that describe some of
these black hole alternatives are similar to some dark matter
candidates, leading to the suggestion that massive compact objects
could be dark matter condensates rather than black holes.

Other recently proposed black hole alternatives are motivated by a
desire to avoid the information paradox --- the loss of information
through the black hole's event horizon.  Such models find ways of
eliminating the event horizon altogether, for instance by replacing
the event horizon with a hard surface surrounding a ball of negative
energy density (the ``gravastar'' model) {\cite{mm01}}, or by
postulating that spacetime itself undergoes a phase transition in the
presence of very strong gravitational fields {\cite{chls}}.  If such
objects exist in nature, they should have a deep, strong field
structure very different from that of black holes {\cite{birkhoff}}.

Astronomical measurements are now becoming able to probe into the very
strong field of compact objects: optical and infrared observations
track stellar orbits at the core of the Milky Way, probing the
spacetime of the presumed black hole at Sgr A* {\cite{ghez,genzel}};
x-ray observations of quasi-periodic oscillations from black hole
candidates carry information about gas in the hole's deep strong field
{\cite{psaltis}}; and future gravitational-wave observations may be
able to track the sequence of orbits followed by a compact body that
slowly spirals into a massive black hole {\cite{annals,bc04}}.  The
question of whether these objects are truly the black holes of general
relativity or are described by some alternative model reduces to the
question of how one may use measurements of orbital properties to map
the spacetime structure (i.e., the gravitational field) of these
objects {\cite{rees04}}.  One thus needs to be able to relate the
properties of the measured orbits to the structure of the central
gravitating objects.  A powerful way of doing this is by a multipole
expansion of the compact object's spacetime.

\subsection{Multipoles of massive compact objects}
\label{subsec:multipoles}

In Newtonian theory, the gravitational field of a body is simply
described by expanding the potential in spherical harmonics.  The
potential $\Phi$ must satisfy
\begin{eqnarray}
\nabla^2\Phi &=& 4\pi G\rho \qquad \mbox{(interior)}\;,
\nonumber\\
&=& 0 \qquad \mbox{(exterior)}\;.
\label{eq:poisson}
\end{eqnarray}
In the exterior, the potential may be expanded as
\begin{equation}
\Phi = -G\sum_{lm} \frac{M_{lm}Y_{lm}}{r^{l+1}}\;.
\label{eq:exterior}
\end{equation}
By matching to an expansion of the interior solution and enforcing
Eq.\ (\ref{eq:poisson}), we see that the coefficients $M_{lm}$ are
{\it mass multipole moments}: numbers that describe the angular
distribution of matter inside the star.  For simplicity, let us focus
for a moment on axially symmetric objects, so that only $m = 0$
matters.  Then, for example, $M_{00} = M$, the total mass of the
object.  By an appropriate choice of the center of our coordinate
system, we put the moment $M_{10} = 0$.  The first interesting moment
is $M_{20}$, the quadrupole moment of the object.  This moment has the
form ${\cal Q} M L^2$, where $L$ is the object's ``size'' (e.g., its
mean radius), and the dimensionless number ${\cal Q}$ describes the
quadrupolar deformation.  Higher moments can likewise be interpreted
as $l$-polar moments of the mass distribution.  Because these moments
directly determine the gravitational potential outside of the
gravitating object, one can measure properties of its mass
distribution by measuring the ``shape'' of the gravitational
potential.  Doing so by studying the properties of satellite orbits is
the science of {\it geodesy}.

Somewhat remarkably, such a description describes the exterior
spacetimes of bodies in general relativity as well.  For any
gravitating body that is stationary, axisymmetric, and reflection
symmetric across the equator --- encompassing black holes plus a wide
variety of perturbations --- the exterior spacetime is fully specified
by a pair of multipole moment families: the mass multipole moment
$M_l$, plus the {\it current} multipole moment $S_l$ {\cite{g70,h74}}.
(Since we have restricted ourselves to axisymmetry, we only consider
$m = 0$.  We henceforth drop this subscript.)  The current moment
$S_l$ does not appear in Newtonian theory; it reflects the fact that
mass and energy flows gravitate in general relativity.  For example,
the moment $S_1$ is the magnitude of the spin angular momentum of the
body.

If the gravitating body is a Kerr black hole, then the values of the
mass and current moments are strongly restricted: in units in which $G
= 1$, $c = 1$ (which we use throughout this paper), we must have
{\cite{h74}}
\begin{equation}
M_l + i S_l = M(i a)^l\;.
\label{eq:kerr_multipoles}
\end{equation}
This equation tells us that $M_0 = M$, the total mass of the Kerr
black hole, and $S_1 = aM$, the magnitude of the black hole's spin
angular momentum in these units --- precisely what we already expect.
More interestingly, {\it all} higher moments are completely determined
by these two values.  The exterior spacetime of a Kerr black hole is
completely determined by its two lowest multipole moments --- its mass
and spin.

This is nothing more than a restatement of the ``no hair'' theorem
{\cite{nohair1,nohair2,nohair3,nohair4}}.  By analogy with geodesy,
this suggests that one can test the no hair theorem by measuring
orbits near black holes.  Using a spacetime that does not necessarily
assume the Kerr form of the moments, one could then determine $M_l$
and $S_l$.  If that object is in fact a black hole as described by
general relativity, the only free moments are those for $l = 0$ and $l
= 1$.  Once they have been determined, {\it all} higher moments are
constrained via Eq.\ (\ref{eq:kerr_multipoles}).  Such ``geodesy for
black holes'' (which has been given the names ``bothrodesy'' and
``holiodesy'' {\cite{odyssey}}) would provide a stringent test of the
black hole nature of massive compact objects in the universe.

The first detailed analysis of how one might be able to falsify the
black hole nature of a massive compact object was by Fintan Ryan
{\cite{r_multi}}.  Ryan showed how to build the spacetime of an object
with arbitrary multipole structure, and then analyzed the orbits of
small bodies in that spacetime.  (``Small'' means that these bodies do
not themselves significantly distort the spacetime, and so can be
treated as following approximately geodesic trajectories).  His
analysis demonstrated that the accumulated orbital phase was sensitive
to these multipoles.  Orbital phase (or some surrogate of this phase)
is directly observed by, for example, x-ray timing (today) and
gravitational-wave detectors (future).  One could thus imagine using
measurements of accumulated orbital phase to test the black hole
nature of a massive compact object.  Focusing on gravitational-wave
measurements with the planned space-based laser interferometer LISA
{\cite{lisa}}, Ryan showed that enough multipoles should be measurable
to easily falsify the black hole hypothesis.  In many cases, enough
multipoles would be measurable (up to $l \sim 5$ or 6) to stringently
constrain the object's black hole nature {\cite{r_measure}}.

Unfortunately, the multipole expansion used by Ryan does not work very
well in the deep strong fields of massive black holes, where one
expects orbital phases to most stringently test the black hole
hypothesis.  Multipole moments essentially label different powers of a
$1/r$ expansion.  In the strong field of a black hole (small $r$),
such an expansion is not going to be very useful {\cite{moderate}}.
The in-utility of this expansion is reflected by the extremely large
number of terms that must be kept to describe a spacetime with
arbitrary multipole moments at small radius (cf.\ Ref.\
{\cite{r_multi}}, Sec.\ III).

\subsection{Bumpy black holes}
\label{subsec:bumpy}

We advocate a different approach.  The reason for introducing a
multipolar expansion is to describe a candidate spacetime differing
from that of a black hole.  If one accepts as a starting point the
idea that the black hole hypothesis {\it probably} describes the
massive compact objects in question, then one just needs a spacetime
to compare against that differs {\it slightly} from that of a black
hole.  Our goal is then to set up a null experiment: we find a trial
spacetime that exhibits slight deviations from the spacetime of a
black hole.  If the black holes of nature are the black holes of
general relativity, we will measure the deviation to be zero.

Past work on candidate objects to test the black hole hypothesis has
focused primarily on boson and soliton stars
{\cite{r_boson,r_measure}}.  Though of great intrinsic interest, there
is no natural way for a boson star spacetime to smoothly limit to the
spacetime of a black hole.  If the massive compact objects we observe
in the universe are in fact black holes, then tests based on the boson
star model will not provide useful constraints on orbit observations.
As a ``straw man'' for the black hole hypothesis, boson stars may
unfortunately contain too much straw.

We advocate instead the use of {\it bumpy black holes}: objects that
have a multipolar structure that is very nearly, but not quite, that
of a black hole.  As the name suggests, these are black holes with
small bumps on them.  If the universe's observed massive compact
objects are in fact black holes, then we will find that the amplitude
of these bumps is zero (within measurement uncertainty).  A bumpy
black hole should be a trial spacetime that behaves well deep into the
strong field, and that exhibits a {\it controllable}\ deviation from
the Kerr solution.  In particular, these trial spacetimes should
reduce to normal black holes when the the deviation is set to zero ---
bumpy black holes become normal black hole when the bumps are removed.
This reduction to normal black holes is a crucial element of using
bumpy black holes as a basis for a null experiment.

A key piece of our guiding philosophy is that the notion of multipoles
is most useful in the weak field of an object.  One shouldn't be too
attached to multipole moments if the goal is an analysis that applies
to strong gravitational fields.  Of course, by taking the weak field
(large $r$) limit of the spacetimes we construct, bumpy black holes
are very usefully described using multipoles.  Indeed, the goal of our
detailed calculations will be to assemble a perturbation that is
purely quadrupolar when examined in the weak field.  Our construction,
however, works very well deep in the strong field, which is crucial
for applying these notions to observations.

An important question to ask at this point is {\it how} one can build
a stationary spacetime corresponding to a bumpy black hole.  A key
portion of the proof of the no hair theorem demonstrates that any
deformation to a black hole will tend to radiate very quickly,
removing the bump and pushing us back to the Kerr black hole solution
{\cite{nohair3,nohair4}}.  Some mechanism must exist to maintain the
bump.  This is likely to require unphysical matter; the example which
we describe in fact requires naked singularities.  One might object
that a bumpy black hole spacetime is thus, by construction,
unphysical.  Our viewpoint is that the physicality of these spacetimes
is {\it irrelevant}.  Our goal in this analysis is {\it not} to build
a spacetime which might conceivably exist in nature.  Instead, we wish
to build a black hole straw man with just the right amount of straw to
probe the nature of massive compact objects.

\subsection{Overview of this paper}
\label{subsec:overview}

The goal of this paper is to present the bumpy black hole concept, to
show how bumpy black hole spacetimes are generated, and to demonstrate
that the magnitude of the bumps is encoded in the accumulated phase of
the hole's orbits.  We focus upon axisymmetric distortions of black
holes --- even in axisymmetry, an incorrect moment is enough to
falsify the black hole hypothesis for a massive compact object.

We have argued that the language of multipoles is not useful for
describing an object's strong field orbits.  To substantiate this
argument, we review the multipole description of spacetimes and their
orbits in Sec.\ {\ref{sec:multipole}}, summarizing the key results of
Ryan {\cite{r_multi}}.  Ryan's formulae and the detailed description
of the spacetime in the multipole language show that, as we try to
characterize the massive object's strong field, the description
becomes extremely complicated.  Although it is possible in principle
to use this description to develop tools for mapping spacetimes, it
does not appear to be the best approach in practice.

We then begin our detailed presentation of bumpy black holes.  In
Sec.\ {\ref{sec:basics}}, we show how to build a bumpy black hole
spacetime.  The spacetime of a stationary, axisymmetric object is
fully described by the Weyl metric {\cite{w18}},
\begin{equation}
ds^2 = -e^{2\psi}dt^2 + e^{2\gamma - 2\psi}(d\rho^2 + dz^2) +
e^{-2\psi}\rho^2 d\phi^2\;.
\label{eq:weyl}
\end{equation}
Our strategy is to pick an exact solution $\psi = \psi_0(\rho,z)$,
$\gamma = \gamma_0(\rho,z)$ for which the line element (\ref{eq:weyl})
describes a black hole.  For this first analysis, we specialize our
background to Schwarzschild black holes; generalization to Kerr black
holes should be straightforward in principle (though it may be
somewhat involved algebraically).  We then use this exact solution as
a background against which to introduce a perturbation, putting $\psi
= \psi_0 + \psi_1$ and requiring $\psi_1/\psi_0 \ll 1$; a similar
perturbation is defined for $\gamma$.  The perturbations are
constrained by the requirement that they satisfy the vacuum Einstein
equations, expanded to first order.

This formulation of the metric is particularly useful because the
function $\psi$ reduces to the Newtonian gravitational potential of
the source in the weak field.  We therefore choose our perturbation
$\psi_1$ in such a manner that the weak-field perturbation can be
thought of as changing the source's multipoles {\it as measured in the
weak field}.  We then solve the linearized Einstein equations in order
to specify the perturbation throughout the exterior spacetime of the
bumpy black hole.

Before specifying our perturbations, we first examine the properties
of orbits in the bumpy hole's equatorial (reflection symmetry) plane
(Sec.\ {\ref{sec:orbits}}).  Many useful quantities can be computed in
terms of the perturbations $\psi_1$ and $\gamma_1$ --- the orbit's
energy $E$, angular momentum $L$, and the location of the last stable
orbit (a separatrix in orbital phase space dividing dynamically stable
and unstable orbits).  We also write down an equation describing the
differential advance of the orbit's periapsis.  The periapsis shift
arises from a mismatch between the radial and azimuthal orbital
frequencies; as such, it can be a sensitive probe of the spacetime.
Deviations of this shift from the canonical Schwarzschild value encode
the black hole's bumpiness.

We choose particular perturbations in Secs.\ {\ref{sec:points}} and
{\ref{sec:ring}}.  A very simple and useful one is that of a point
mass near the black hole.  We build a bumpy black hole in Sec.\
{\ref{sec:points}} by placing a pair of point masses with mass $\mu/2$
each near the hole's ``north'' and ``south'' poles.  The same system
was used by Suen, Price, and Redmount (SPR) {\cite{spr88}} to set up
an analysis of a black hole with a deformed event horizon.  Our
analysis is similar to that of SPR, though we do not focus on the
region of spacetime near the horizon.  We build a second type of bumpy
black hole in Sec.\ {\ref{sec:ring}} by placing a ring of mass $\mu$
about the hole's equator.

As the analysis of Secs.\ {\ref{sec:points}} and {\ref{sec:ring}}
shows, both the polar point mass and the equatorial ring do indeed
change the metric's quadrupole moment.  We demonstrate this by
calculating weak-field periapsis precession in these spacetimes and
showing that the shift contains a term which is exactly what we expect
for weak-field quadrupolar distortions (computed in Appendix
{\ref{sec:newton}}).  Unfortunately, these perturbations also change
the metric's {\it monopole}\ moment (i.e., its mass).  Fortunately, we
can build a purely quadrupolar distortion by combining {\it negative}
mass polar points with a positive mass equatorial ring, or vice versa.
(Bearing in mind that our goal is to build trial spacetimes for
testing the black hole hypothesis, the unphysicality of a negative
perturbing mass is not a concern.)  In Sec.\ {\ref{subsec:both_weak}},
we show that the weak-field periapsis precession with this combined
mass distribution is identical to that of a Schwarzschild black hole
plus a quadrupolar deformation.  The points + ring perturbation to a
Schwarzschild black hole thus perfectly matches our requirements for a
bumpy black hole.  We investigate the strong-field character of this
spacetime in Sec.\ {\ref{subsec:both_strong}}, showing in particular
that the hole's bumpiness is usefully encoded in the strong-field
periapsis precession.

Concluding discussion is given in Sec.\ {\ref{sec:concl}}.  In
particular, we outline further work that should be done to connect the
bumpy black hole concept to future astrophysical observations.  Chief
among the tasks needed is a generalization to Kerr black hole
backgrounds; some steps in this direction are outlined in Appendix
{\ref{sec:kerr}}.

\section{Metrics and multipole moments: an overview}
\label{sec:multipole}

As stated in the Introduction, one can build a spacetime by specifying
a set of mass and current multipole moments $(M_l,S_l)$.  In
actuality, one builds a spacetime from a set of coefficients $a_{jk}$
which determine orbital characteristics; the multipole moments can
then be extracted from these coefficients.  In this section we briefly
describe the details of this analysis, and discuss why we believe this
is not the most effective way to map black hole spacetimes in
practice.  Much of our presentation is essentially a synopsis of
Fintan Ryan's calculation; see Ref.\ {\cite{r_multi}} for detailed
discussion.

To begin, we must pick a spacetime sufficiently general to encompass
the stationary, axisymmetric sources we wish to describe.  Ryan begins
with the line element
\begin{equation}
ds^2 = -F\left(dt - \omega\,d\phi\right)^2 +
\frac{1}{F} \left[e^{2\gamma}\left(d\rho^2 +
dz^2\right) + \rho^2 d\phi^2\right]\;.
\label{eq:ryan_metric}
\end{equation}
The functions $F$, $\gamma$, and $\omega$ depend on $\rho$ and $|z|$.
The radial coordinate $\rho$ labels displacement from the source's
symmetry axis; $z$ labels displacement above or below the source's
``equatorial'' plane.  By construction, the spacetime is stationary
and axisymmetric ($F$, $\gamma$, and $\omega$ do not depend on $t$ or
$\phi$), and is reflection symmetric about the equatorial plane
(dependence on $|z|$).

For an axial and reflection symmetric spacetime, the metric functions
$F$, $\gamma$, and $\omega$ can be generated from the {\it Ernst
potential} ${\cal E}$.  This function and a related complex function
$\tilde\xi$ are defined via
\begin{equation}
{\cal E} = F + i\Psi = \frac{\sqrt{\rho^2 + z^2} - \tilde\xi}
{\sqrt{\rho^2 + z^2} + \tilde\xi}\;,
\label{eq:ernst}
\end{equation}
where the function $\Psi$ is related to $\omega$ by
\begin{equation}
\omega(\rho,z) = -\int_\rho^\infty
\left(\frac{\rho^\prime}{F^2}\frac{\partial\Psi}{\partial z}\right)
d\rho^\prime\;.
\end{equation}
The quantity in parentheses under the integral is evaluated at
constant $z$.  The function $\gamma$ can be determined once $F$ and
$\omega$ are known: the vacuum Einstein equations tell us
\begin{eqnarray}
\frac{\partial\gamma}{\partial\rho} &=&
\frac{1}{4}\frac{\rho}{F^2}
\left[\left(\frac{\partial F}{\partial\rho}\right)^2 -
\left(\frac{\partial F}{\partial z}\right)^2\right] -
\frac{1}{4}\frac{F^2}{\rho}
\left[\left(\frac{\partial\omega}{\partial\rho}\right)^2 -
\left(\frac{\partial\omega}{\partial z}\right)^2\right]\;,
\label{eq:dgamdrho0}
\\
\frac{\partial\gamma}{\partial z} &=&
\frac{1}{2}\frac{\rho}{F^2}
\frac{\partial F}{\partial\rho} \frac{\partial F}{\partial z} -
\frac{1}{2}\frac{F^2}{\rho}
\frac{\partial\omega}{\partial\rho}
\frac{\partial\omega}{\partial z}\;,
\label{eq:dgamdz0}
\end{eqnarray}
Then, $\gamma$ can be found at any point in spacetime with an
appropriate line integral.  See Ref.\ {\cite{wald}}, pp.\ 165 -- 167
for further discussion.  For our purposes, the main thing to note is
that knowledge of ${\cal E}$ --- or equivalently $\tilde\xi$ ---
specifies the {\it entire} spacetime metric.

The function $\tilde\xi$ can be expanded as
\begin{equation}
{\tilde\xi} = \sum_{j,k=0}^\infty a_{jk} \frac{\rho^j z^k}
{\left(\rho^2+z^2\right)^{j+k}}\;.
\label{eq:xi}
\end{equation}
The index $j$ is strictly even.  If $k$ is even, $a_{jk}$ is real; if
$k$ is odd, $a_{jk}$ is imaginary.  From these coefficients, it is
relatively straightforward to extract the spacetime's multipole
moments $(M_l,S_l)$ using an algorithm developed by Fodor,
Hoenselaers, and Perj\'es (FHP) {\cite{fhp89}}.  By a recursive
procedure that involves repeatedly differentiating $\tilde\xi$ and
gathering terms, FHP show that $a_{jk}$ can be written in terms of the
mass and current moments [see Ref.\ {\cite{r_multi}}, Eqs.\ (35) --
(41)].  For example, one can show that
\begin{eqnarray}
a_{0l} &=& M_l + i S_l + {\rm LOM}\;,
\label{eq:a_0l}\\
a_{l0} &=& (-1)^{l/2}\frac{(l-1)!!}{l!!} M_l + {\rm LOM}\;,
\label{eq:a_l0}\\
a_{l-1,1} &=& i(-1)^{(l-1)/2}\frac{l!!}{(l-1)!!} S_l + {\rm LOM}\;.
\label{eq:a_l-11}
\end{eqnarray}
``LOM'' is an abbreviation for ``Lower Order Moments'': a complicated
(but known) sum of various combinations of $M_j$ and $S_k$ with $j <
l$ and $k < l$.

Ryan {\cite{r_multi}} uses the coefficients $a_{jk}$ and their
relationship to $(M_l,S_l)$ as the basis for his spacetime mapping
procedure.  Since the spacetime is fully determined by $a_{jk}$, it
follows that its orbits are likewise determined.  An orbit in the
spacetime (\ref{eq:ryan_metric}) is governed by three orbital
frequencies: $\Omega_\phi$, related to the time to cover $2\pi$
radians of azimuth; $\Omega_\rho$, characterizing oscillations in the
$\rho$ coordinate; and $\Omega_z$, for oscillations in the $z$
coordinate.  Ryan shows (via a power law expansion using orbital speed
as an expansion parameter) how these frequencies depend on $a_{jk}$.
Measurement of $(\Omega_\phi, \Omega_\rho, \Omega_z)$ (or accumulated
phases associated with these frequencies) can therefore be used to
measure $a_{jk}$.  Ryan then inserts the measured values of $a_{jk}$
into the FHP algorithm, determining the multipole moments of the
spacetime.  Using this procedure, Ryan showed that future
gravitational-wave measurements with LISA should be able to determine
enough moments to strongly constrain the black hole nature of massive
compact objects {\cite{r_measure}}.

Though sufficient to prove the principle, we believe that this
procedure is not useful in practice for mapping the spacetimes of
objects believed to be black holes.  Referring to Eq.\ (\ref{eq:xi}),
we see that the coefficients $a_{jk}$ are essentially labels for an
expansion in inverse distance.  For strong field orbits
($\rho,z\sim{\rm a\ few} \times M$), a large number of these
coefficients must be kept in order to model the spacetime with
sufficient accuracy.  One might hope that the coefficients $a_{jk}$
become small for large $j$ and $k$, making it possible to truncate the
expansion of $\tilde\xi$.  This is not the case: because of the
coupling to lower order moments [cf.\ Eqs. (\ref{eq:a_0l}) --
(\ref{eq:a_l-11})], these coefficients generically remain large even
if the body has only a small number of non-zero multipole moments.

In this language, the description of the spacetime and hence of its
orbits becomes extremely complicated in the strong field.  This makes
testing whether a spacetime is close to that of a black hole very
difficult.  Naively, one might imagine requiring that a spacetime have
the multipole moment structure
\begin{equation}
M_l + i S_{l} = M(i a)^l + \delta M_l + i \delta S_l\;,
\label{eq:kerr_distort}
\end{equation}
and then developing a formalism similar to that described here to
place observational limits on the deviations $(\delta M_l,\delta
S_l)$.  (Indeed, this how we originally conceived of this analysis.)
One quickly discovers that the algebraic complexity associated with
this approach is immense.  Though no doubt possible in principle, a
multipolar prescription like Eq.\ (\ref{eq:kerr_distort}) does not
easily translate into a {\it practical} scheme for constraining the
properties of massive compact objects.

The lesson appears to be that multipoles, though conceptually clean
and offering a beautiful description of weak-field gravity, simply
aren't the best tools in the strong field.  This shouldn't be
surprising in a nonlinear theory like general relativity: since
multipoles are basically labels in an inverse distance expansion (as
we have repeatedly emphasized), a description that is clean in the
weak field can easily turn into a mess when the nonlinearities are
important.

\section{Building the spacetime of a bumpy black hole}
\label{sec:basics}

Keeping in mind that the goal is just to build some candidate
spacetime to be used as a straw man in testing the black hole
hypothesis, we advocate a different approach.  Our goal is to develop
a family of spacetimes corresponding to stationary perturbed black
holes: bumpy black holes.  These spacetimes include black holes as a
subset --- we simply set the magnitude of the perturbation to zero.
We construct these spacetimes in a manner that makes an exploration of
its strong-field properties simple.

We will focus on axisymmetric spacetimes, in keeping with our goal of
analyzing axisymmetric deformations of Kerr black holes.  For this
first analysis, we will further specialize to deformations of
Schwarzschild black holes.  Stationary, axisymmetric deformations of
Schwarzschild black holes were in fact studied by SPR {\cite{spr88}}
with the goal of characterizing distortions to the event horizon.
(Their analysis was a part of the ``Membrane Paradigm''; see Ref.\
{\cite{membrane}}.)  Their calculation makes an ideal starting point
for our analysis; the following discussion closely follows Ref.\
{\cite{spr88}}, Sec.\ IIIA.

As mentioned in the Introduction, the spacetimes we consider are
described by the Weyl metric {\cite{w18}}:
\begin{equation}
ds^2 = -e^{2\psi}dt^2 + e^{2\gamma - 2\psi}\left(d\rho^2 + dz^2\right)
+ e^{-2\psi}\rho^2 d\phi^2\; .
\label{eq:weyl_cyl}
\end{equation}
The vacuum Einstein equations for this line element reduce to
\begin{eqnarray}
\frac{\partial^2\psi}{\partial\rho^2} &+& \frac{1}{\rho}
\frac{\partial\psi}{\partial\rho} + \frac{\partial^2\psi}{\partial
z^2} = 0\;,
\label{eq:d2psidrho2}\\
\frac{\partial\gamma}{\partial\rho} &=&
\rho\left[\left(\frac{\partial\psi}{\partial\rho}\right)^2 -
\left(\frac{\partial\psi}{\partial z}\right)^2\right]\;,
\label{eq:dgamdrho}\\
\frac{\partial\gamma}{\partial z} &=&
2\rho\frac{\partial\psi}{\partial\rho}\frac{\partial\psi}{\partial z}
\;.
\label{eq:dgamdz}
\end{eqnarray}
Equations (\ref{eq:dgamdrho}) and (\ref{eq:dgamdz}) are identical to
Eqs.\ (\ref{eq:dgamdrho0}) and (\ref{eq:dgamdz0}) with $\omega = 0$
and $F = e^{2\psi}$.  Note that Eq.\ (\ref{eq:d2psidrho2}) implies
$\psi$ is a harmonic function in a fictitious Euclidean space with
cylindrical coordinates $\rho$, $z$, and $\phi$ {\cite{spr88,wald}}.

Following SPR {\cite{spr88}}, we observe that the function $\psi$ is
the Newtonian potential far from a source.  A reasonable way to
perturb $\psi$ is by adding potential terms that correspond to
particular mass distributions perturbing the background black hole;
distributions that change the weak-field multipole structure of the
hole are particularly interesting.  [Note the linearity of Eq.\
(\ref{eq:d2psidrho2}) --- {\it exact} solutions for $\psi$ can be
constructed by superposition.]  We only require that the mass $\mu$ of
the perturbations be small compared to the mass $M$ of the black hole,
allowing us to expand $\psi$ and $\gamma$ to first order.

Notice that, with our form of the perturbation, we expand in powers of
$\mu/M$, rather than $1/r$ (where $r$ is some measure of distance from
the source).  The approximation we introduce should thus be
well-behaved for any $r$, including into the strong field.  Notice
also that, because Eqs.\ (\ref{eq:d2psidrho2}) -- (\ref{eq:dgamdz})
come from the {\it vacuum} Einstein equations, our metric will only
hold where there is no matter.  To perturb the background black hole,
we will add matter in the form of point particles and one-dimensional
rings.  The metric will therefore not apply at the points containing
that matter; it will in fact be singular at those locations.  As long
as we only examine regions of spacetime external to these perturbing
sources, this singular behavior does not pose any difficulty.

For describing black holes, it is convenient to use prolate spheroidal
coordinates $u$ and $v$:
\begin{eqnarray}
\rho &=& M\sinh u\sin v\;,
\label{eq:rho}\\
z &=& M\cosh u\cos v\;.
\label{eq:z}
\end{eqnarray}
The coordinate $v \in [0,\pi]$ is a polar angle; $u\in [0,\infty)$ is
effectively a radial coordinate.  These coordinates cover the entire
{\it exterior} Schwarzschild spacetime: the coordinate $u = 0$ maps to
the event horizon, $r = 2M$ [cf.\ Eq.\ (\ref{eq:schw_coords}) below].
In the Weyl coordinates ($\rho,z$), this corresponds to a cylindrical
rod at $\rho = 0$ running from $z = -M$ to $z = M$.

The line element (\ref{eq:weyl_cyl}) becomes
\begin{equation}
ds^2 = -e^{2\psi}dt^2 + M^2e^{2\gamma - 2 \psi}(\sinh^2 u + \sin^2 v)
(du^2 + dv^2) + M^2e^{-2\psi}\sinh^2 u\sin^2v\,d\phi^2 \;.
\label{eq:weyl_prolate}
\end{equation}
We now put $\psi = \psi_0 + \psi_1$, $\gamma = \gamma_0 + \gamma_1$.
We take the perturbations $(\psi_1,\gamma_1)$ to be small compared to
$(\psi_0,\gamma_0)$:
\begin{equation}
\psi_1 \sim \frac{\mu}{M}\psi_0\;,
\end{equation}
and likewise for $\gamma$.  The Einstein equations constraining
$\gamma$, (\ref{eq:dgamdrho}) and (\ref{eq:dgamdz}), become
\begin{eqnarray}
\frac{\partial\gamma_1}{\partial u} &=&
2\rho^2\cot v\left(
\frac{\partial\psi_0}{\partial\rho}\frac{\partial\psi_1}{\partial z} +
\frac{\partial\psi_0}{\partial z}\frac{\partial\psi_1}{\partial\rho}
\right) + 2\rho z\tan v\left(
\frac{\partial\psi_0}{\partial\rho}\frac{\partial\psi_1}{\partial\rho} -
\frac{\partial\psi_0}{\partial z}\frac{\partial\psi_1}{\partial z}
\right)\;,
\label{eq:dgam1du}\\
\nonumber\\
\frac{\partial\gamma_1}{\partial v} &=&
2\rho^2\cot v\left(
\frac{\partial\psi_0}{\partial\rho}\frac{\partial\psi_1}{\partial\rho} -
\frac{\partial\psi_0}{\partial z}\frac{\partial\psi_1}{\partial z}
\right) + 2\rho z\tan v\left(
\frac{\partial\psi_0}{\partial\rho}\frac{\partial\psi_1}{\partial z} +
\frac{\partial\psi_0}{\partial z}\frac{\partial\psi_1}{\partial\rho}
\right)\;.
\label{eq:dgam1dv}
\end{eqnarray}
We have only kept terms to leading order in the perturbation.  It
turns out that these two equations will {\it overdetermine} the
solution (cf.\ {\cite{wald}}, p. 167 for discussion).  For our
purposes, this means that we need integrate only one of them to
determine $\gamma_1$ (up to a constant of integration).  We will use
Eq.\ (\ref{eq:dgam1dv}); the solution we find also satisfies
(\ref{eq:dgam1du}), as is easily verified by direct substitution.

At this point, we specialize our background to the Schwarzschild
metric: we put {\cite{spr88}}
\begin{eqnarray}
\psi_0 &=& \ln\tanh u/2\;,
\label{eq:psi0}\\
\gamma_0 &=& -\frac{1}{2}\ln\left(1+\frac{\sin^2 v}{\sinh^2
u}\right)\;.
\label{eq:gam0}
\end{eqnarray}
Using Eqs.\ (\ref{eq:psi0}) and (\ref{eq:gam0}) and changing all
instances of $\rho,z$ to $u,v$ as appropriate, Eq.\
(\ref{eq:dgam1dv}) becomes
\begin{equation}
\frac{\partial\gamma_1}{\partial v} =
\frac{2\left[\tan v \left(\partial\psi_1/\partial v\right) -
             \tanh u \left(\partial\psi_1/\partial u\right)\right]}
{\sinh u\left(\coth u\tan v+\tanh u\cot v\right)}\;.
\label{eq:dgam1dv_fin}
\end{equation}
For completeness, the equation for $\partial\gamma_1/\partial u$
becomes
\begin{equation}
\frac{\partial\gamma_1}{\partial v} = \frac{2\left[\cot v
\left(\partial\psi_1/\partial u\right) + \tanh u
\left(\partial\psi_1/\partial v\right)\right]}{\sinh u \left(\coth
u\tan v + \tanh u\cot v\right)}\;;
\label{eq:dgam1du_fin}
\end{equation}
as already discussed, we will only focus on Eq.\
(\ref{eq:dgam1dv_fin}).

To solve for $\gamma_1$, we will first {\it impose} a particular
$\psi_1$, taking advantage of the fact that $\psi_1$ can be thought of
as a perturbation to the distant, Newtonian gravitational field of the
black hole.  We then integrate $\partial\gamma_1/\partial v$ with
respect to $v$ to find $\gamma_1(u,v)$, being careful to choose an
appropriate integration constant by demanding that the perturbation
vanish far from the black hole.  The $\gamma_1(u,v)$ we find then
automatically satisfies Eq.\ (\ref{eq:dgam1du}) or
(\ref{eq:dgam1du_fin}).

In preparation for these calculations, it is useful to put this metric
into Schwarzschild-like form.  Plugging Eqs.\ (\ref{eq:psi0}) and
(\ref{eq:gam0}) into Eq.\ (\ref{eq:weyl_prolate}), and making the
coordinate transformation {\cite{spr88}}
\begin{equation}
r = 2M\cosh^2 u/2, \qquad \theta = v\;
\label{eq:schw_coords}
\end{equation}
yields
\begin{equation}
ds^2 = -e^{2\psi_1}\left(1 - \frac{2M}{r}\right)dt^2
+ e^{2\gamma_1 - 2\psi_1}\left(1 - \frac{2M}{r}\right)^{-1}dr^2
+ r^2 e^{2\gamma_1 - 2\psi_1}d\theta^2
+ r^2\sin^2\theta e^{-2\psi_1}d\phi^2\;.
\label{eq:pert_schw}
\end{equation}
Note that the transformation (\ref{eq:schw_coords}) implies
\begin{eqnarray}
\rho &=& r\sin\theta\sqrt{1 - \frac{2M}{r}}\;,
\label{eq:rho_of_r}\\
z &=& \left(r - M\right)\cos\theta\;.
\label{eq:z_of_r}
\end{eqnarray}
These relations can be particularly helpful when describing the
perturbations in Schwarzschild coordinates.

Although the line element (\ref{eq:pert_schw}) is technically exact,
in the following analysis we will only work to first order in the
perturbation.  We thus should really expand $e^{2\psi_1} \simeq 1 +
2\psi_1$, and likewise for $\gamma_1$.  For compactness of notation,
we leave these perturbations in the exponentials, with the caveat that
they must be expanded to first order in all calculations.


Before examining some interesting perturbations, it is useful to
carefully study orbits in this general perturbed spacetime.  Several
results can be found in terms of $\psi_1$ and $\gamma_1$, which
facilitates later analysis.  In the next section, we will examine the
properties of equatorial orbits in this metric, and calculate the rate
of periapse precession.

\section{Equatorial orbits of bumpy black holes}
\label{sec:orbits}

In this section, we study, as much as is possible, properties of bumpy
black hole orbits that do not require specifying the perturbation.  It
is easy to find the geodesics in the equatorial plane ($\theta =
\pi/2$); for this first analysis, we will focus on this simple case.
It is not, in fact, difficult to generalize the equations of motion to
orbits beyond the equatorial plane.  To solve these equations,
however, appears challenging: the equations for $r$ and $\theta$ do
not appear to separate in general (as they do, for example, in the
Kerr case).  We discuss this issue further in Sec.\ {\ref{sec:concl}}.

As discussed in the Introduction, our goal is to understand how black
hole bumpiness is imprinted upon measurable quantities --- in
particular, accumulated phases related to harmonics of the orbital
frequencies.  A simple (and historically important) effect is the
shift of an orbit's periapsis.  This shift is related to the mismatch
between the $r$ and $\phi$ frequencies.  In the weak-field limit, it
describes perihelion precession, well-known from studies of planetary
orbits in our own solar system.  In this section, we will derive a
differential formula for the periapsis shift.

We begin by writing down a Lagrangian for orbiting bodies in this
spacetime: put ${\cal L} = g_{ab} {\dot x}^a {\dot x}^b$ (where
overdot denotes $d/d\tau$, with $\tau$ proper time along an orbit).
Note that ${\cal L} = -1$.  Since our focus here is equatorial orbits,
we further put $\theta = \pi/2$ and $\dot\theta = 0$.  (Thanks to the
reflection symmetry, we are guaranteed that an equatorial trajectory
remains equatorial.)  The result is
\begin{equation}
{\cal L} = -e^{2\psi_1}\left(1 - \frac{2M}{r}\right)\dot{t}^2
+ e^{2\gamma_1 - 2\psi_1}\left(1 - \frac{2M}{r}\right)^{-1}\dot{r}^2
+ e^{-2\psi_1}r^2 \dot{\phi}^2\;.
\label{eq:lagrangian}
\end{equation}
Varying ${\cal L}$ with respect to $t$ yields
\begin{equation}
\frac{d}{d\tau}\left[e^{2\psi_1}\left(1 -
\frac{2M}{r}\right)\dot{t}\right] = 0\;,
\end{equation}
leading us to identify the orbital energy per unit rest mass $\mu$
\begin{equation}
E = e^{2\psi_1}\left(1-\frac{2M}{r}\right)\dot{t}
\label{eq:orb_E_gen}
\end{equation}
as a constant of motion.  By varying with respect to $\phi$, we
likewise identify the orbital angular momentum orthogonal to the
equatorial plane per unit $\mu$:
\begin{equation}
L = e^{-2\psi_1}r^2\dot{\phi}\;.
\label{eq:orb_L_gen}
\end{equation}

An equation for the radial motion follows from ${\cal L} = -1$:
\begin{equation}
\dot{r}^2 = e^{-2\gamma_1}
\left[E^2 - e^{2\psi_1}\left(1 - \frac{2M}{r}\right)
\left(1 + \frac{e^{2\psi_1}L^2}{r^2}\right)\right]\;.
\label{eq:rdot_gen}
\end{equation}
It is easy to see that this reduces to the usual Schwarzschild
equation of motion for $\psi_1 = \gamma_1 = 0$ [compare, e.g., Ref.\
{\cite{mtw}, Eq. (25.16a)].  For the calculations we will perform
momentarily, it is useful to multiply this by $r^3$, defining
\begin{eqnarray}
R(r) &=& r^3\dot{r}^2
\nonumber\\
&=& e^{-2\gamma_1}
\left[(E^2 - e^{2\psi_1})r^3 + 2e^{2\psi_1}M r^2
  - e^{4\psi_1} L^2 r + 2 e^{4\psi_1} L^2M\right]\;.
\label{eq:Rdef}
\end{eqnarray}
Using $R(r)$, it is straightforward to remap the constants $E$ and $L$
to parameters that directly characterize the orbit.  It is helpful to
reparameterize $r$,
\begin{equation}
r = \frac{p}{1 + \varepsilon\cos\chi}\;.
\label{eq:r_remap}
\end{equation}
As $\chi$ oscillates from $0$ to $\pi$ to $2\pi$, the orbiting body
moves from periapsis, $r_p = p/(1 + \varepsilon)$, to apoapsis, $r_a =
p/(1 - \varepsilon)$, and back.  In Newtonian gravity, this
reparameterization facilitates identifying the orbits as closed
ellipses; $p$ is the orbit's semi-latus rectum and $\varepsilon$ its
eccentricity.  This intuition remains very useful in general
relativity, though the ellipses do not close --- the time for $\chi$
to go from $0$ to $2\pi$ is greater than the time for the orbit to
move through $2\pi$ radians of azimuth $\phi$.

The periapse and apoapse radii are the orbit's turning points.  By
definition, this means that $\dot{r}$ and hence $R(r)$ equal zero at
these points.  This makes it possible to find $E$ and $L$ as functions
of $p$ and $\varepsilon$: simultaneously solving $R(r_p) = 0$ and
$R(r_a) = 0$ yields
\begin{equation}
E(p,\varepsilon) = e^{[\psi_1(r_p) + \psi_1(r_a)]}\sqrt{\frac{
\left[(1 + \varepsilon)^2 e^{2\psi_1(r_p)}
-(1 - \varepsilon)^2 e^{2\psi_1(r_a)}\right]
\left[p^2 - 4 M p + 4 M^2(1 - \varepsilon^2)\right]}
{p\left\{
e^{4\psi(r_p)}(1 + \varepsilon)^2
\left[p - 2 M(1 + \varepsilon)\right] -
e^{4\psi(r_a)}(1 - \varepsilon)^2
\left[p - 2 M(1 - \varepsilon)\right]
\right\}}}\;,
\label{eq:en_of_p_e}
\end{equation}
\begin{equation}
L(p,\varepsilon) = p\sqrt{\frac{
e^{2\psi(r_p)}\left[p - 2 M(1 + \varepsilon)\right] -
e^{2\psi(r_a)}\left[p - 2 M(1 - \varepsilon)\right]}
{e^{4\psi(r_p)}(1 + \varepsilon)^2
\left[p - 2 M(1 + \varepsilon)\right] -
e^{4\psi(r_a)}(1 - \varepsilon)^2
\left[p - 2 M(1 - \varepsilon)\right]}}\;.
\label{eq:angmom_of_p_e}
\end{equation}
For orbits that are circular ($\varepsilon = 0$), $r_p = r_a$, so the
conditions $R(r_p) = 0$, $R(r_a) = 0$ do not provide separate
information.  In this case, we require $\partial R/\partial r = 0$;
this condition guarantees that $\dot r$ remains zero.  Solving $R =
0$, $\partial R/\partial r = 0$ yields
\begin{equation}
E(p,\varepsilon = 0) =
e^{\psi_1(p)}\sqrt{
\left(1 - \frac{2M}{p}\right)
\left[\frac
{p - 2 M - p(p - 2 M){\partial\psi_1/\partial r}}
{p - 3 M - 2p(p - 2 M){\partial\psi_1/\partial r}}\right]}\;,
\label{eq:en_circ}
\end{equation}
\begin{equation}
L(p,\varepsilon = 0) =
p e^{-\psi_1(p)}\sqrt{\frac
{M + p(p - 2 M){\partial\psi_1/\partial r}}
{p - 3 M - 2p(p - 2 M){\partial\psi_1/\partial r}}}\;.
\label{eq:angmom_circ}
\end{equation}
When the perturbation $\psi_1$ is set to zero, these expressions
correctly reduce to formulas characterizing $E$ and $L$ for
Schwarzschild black holes:
\begin{eqnarray}
E(p,\varepsilon) &=& \sqrt{\frac
{p^2 - 4 M p + 4 M^2 (1 - \varepsilon^2)}
{p\left[p - M(3 + \varepsilon^2)\right]}}
\\
&=& \frac{1 - 2M/p}{\sqrt{1 - 3M/p}}
\qquad(\varepsilon = 0)\;;
\end{eqnarray}
\begin{eqnarray}
L(p,\varepsilon) &=& p\sqrt{\frac{M}{p - M(3 + \varepsilon^2)}}
\\
&=& p\sqrt{\frac{M}{p - 3M}}
\qquad(\varepsilon = 0)\;.
\end{eqnarray}

The function $R(r)$ also determines the parameters of the {\it last
stable orbit}: solving the equation
\begin{equation}
\frac{\partial R}{\partial r} = 0
\label{eq:LSO_cond}
\end{equation}
with $E = E(p,\varepsilon)$, $L = L(p,\varepsilon)$ determines a
separatrix $p(\varepsilon)$ such that, at fixed eccentricity
$\varepsilon$, orbits with $p < p(\varepsilon)$ are dynamically
unstable.  (For circular orbits, we must solve $\partial^2 R/\partial
r^2 = 0$.)  The result is rather messy, so we do not present it
explicitly.  When the perturbation is turned off, the separatrix is
simply given by $p = (6 + 2\varepsilon)M$.

We are now ready to derive a general formula for bumpy black hole
periapsis precession.  First, note that, via Eq.\ (\ref{eq:r_remap}),
the rate at which the orbital radius changes with respect to $\chi$ is
\begin{equation}
\frac{dr}{d\chi} = \frac{p\varepsilon\sin\chi}{(1 +
\varepsilon\cos\chi)^2} = \frac{r^2\varepsilon\sin\chi}{p}\;.
\label{eq:drdpsi}
\end{equation}
Combining Eqs.\ (\ref{eq:orb_L_gen}), (\ref{eq:rdot_gen}), and
(\ref{eq:drdpsi}), we find
\begin{eqnarray}
\frac{d\phi}{d\chi} &=& \frac{d\phi}{d\tau} \frac{d\tau}{dr}
\frac{dr}{d\chi}
\nonumber\\
&=& e^{2\psi_1[r(\chi)]}e^{\gamma_1[r(\chi)]}
L(p,\varepsilon)\varepsilon\sin\chi p^{-1} \left[E(p,\varepsilon)^2
- e^{2\psi_1[r(\chi)]}\left[1 - \frac{2M}{r(\chi)}\right] \left[1 +
\frac{e^{2\psi_1[r(\chi)]}L(p,\varepsilon)^2}{r(\chi)^2}\right]
\right]^{-1/2}\;.
\label{eq:diff_periapse}
\end{eqnarray}
This equation expresses, in differential form, the amount of azimuth
that accumulates per unit $\chi$.  Integrating $d\phi/d\chi$ over
$0\le\chi\le2\pi$ gives the total accumulated azimuth in one full
orbit.  In Newtonian gravity, this number is $2\pi$ radians ---
Newtonian orbits are closed ellipses.  The periapsis precession is the
amount of ``extra'' azimuth that accumulates in one orbit:
\begin{equation}
\Delta\phi = \int_0^{2\pi} d\chi\frac{d\phi}{d\chi} - 2\pi\;.
\label{eq:periaps_prec}
\end{equation}
Having derived about as many orbital properties as we can in general,
much of the rest of this paper will be devoted to integrating this
equation with particular choices of the perturbing functions $\psi_1$
and $\gamma_1$.  We turn now to a calculation of these perturbations
and an exploration of their effects.

\section{Metric Perturbations I: Point masses at the poles}
\label{sec:points}

In this and the following section, we calculate the first-order metric
perturbation in two special cases: a pair of point masses at the poles
(this section), and a ring of mass about the equator (following
section).  These two perturbations are particularly interesting
because they produce a quadrupolar distortion of the spacetime.

Point masses at the poles were analyzed by SPR in Ref. {\cite{spr88}}.
Their particular focus was on the perturbation at or near the hole's
event horizon; we generalize their result to find the perturbation
throughout the exterior spacetime.

We begin by choosing, in Weyl coordinates $(t,\rho,z,\phi)$, the
$\psi$ perturbation
\begin{equation}
\psi_1^{\rm NP} = -\frac{\mu/2}{\sqrt{\rho^2 + (z - b)^2}}\;.
\label{eq:psi1_np}
\end{equation}
As described previously, this choice has a simple interpretation:
adding a point mass to the system changes the Newtonian potential at
infinity --- $\psi$ --- by that of a point mass.  Equation
(\ref{eq:psi1_np}) describes a point with mass (as measured at
infinity) $\mu/2$ at $z = b$, near the ``north'' pole.  The
corresponding $\psi_1$ for a perturbing mass near the ``south'' pole
is
\begin{equation}
\psi_1^{\rm SP} = -\frac{\mu/2}{\sqrt{\rho^2 + (z + b)^2}}\;.
\label{eq:psi1_sp}
\end{equation}
The complete $\psi$ perturbation is given by adding the contributions
from the north and south poles.

Our choice for $\psi_1$ has a curious property: it corresponds to a
nonspherical naked singularity at $\rho = 0$, $z = \pm b$ (the Curzon
solution {\cite{curzon}}).  One might object that a spacetime which
includes such a singularity can not be physical, since naked
singularities are presumed, by cosmic censorship, not to exist in
nature.  Therefore, perhaps we should exclude such spacetimes from
consideration.

As we have repeatedly emphasized, given the goals of this analysis,
the physicality of our trial spacetime is irrelevant.  We seek a
family of spacetimes that we can compare with those found in nature.
In particular, we ned a family of spacetimes that deviate controllably
and only slightly from black holes.  For the purposes of uncovering
whether a spacetime is a black hole or not, it makes no difference
whether the deviations come from physics we expect or not.  We merely
want to set limits on the extent of any deviations that might exist.

Having specified $\psi_1$, we turn to the computation of $\gamma_1$.
Focus first on the perturbation at the north pole.  Switching to the
prolate, spheroidal coordinates $u$ and $v$, we rewrite
\begin{equation}
\psi_1^{\rm NP} = -\frac{\mu/2} {\left[M^2\sinh^2 u\sin^2 v + (b -
M\cosh u\cos v)^2\right]^{1/2}}\;.
\label{eq:psi1_np_sph}
\end{equation}
Insert this into Eq.\ (\ref{eq:dgam1dv_fin}):
\begin{eqnarray}
\frac{\partial\gamma_1^{\rm NP}}{\partial v} &=& \frac{\mu M \left(b-
M\cosh u\cos v\right)\sin v} {\left[M^2\sinh^2u\sin^2v + (b - M\cosh
u\cos v)^2\right]^{3/2}}\;,
\nonumber\\ &=& \frac{\mu}{M}\frac{(\hat b- \cosh u\cos v)\sin v}
{\left[\sinh^2u\sin^2v + (\hat b - \cosh u\cos v)^2\right]^{3/2}}\;.
\end{eqnarray}
We have defined $\hat b = b/M$.

We now integrate this with respect to $v$ to compute $\gamma_1^{\rm
NP}$.  Several manipulations help.  First, simplify the denominator:
\begin{eqnarray}
\sinh^2 u \sin^2 v + (\hat b - \cosh u\cos v)^2
&=& \sinh^2 u + \hat b^2 -2\hat b\cosh u\cos v + \cos^2 v
\nonumber\\
&=& \left(\cos v - \hat b\cosh u\right)^2 + \hat b^2 - \hat b^2\cosh^2
u + \sinh^2 u
\nonumber\\
&=& \left(\cos v - \hat b\cosh u\right)^2 + \left(1 - \hat
b^2\right)\sinh^2 u\;.
\end{eqnarray}
Next, put $x = \cos v$, so $dx = -\sin v\,dv$:
\begin{equation}
\gamma_1^{\rm NP} = \frac{\mu}{M}
\int\frac{\left(-\hat b + x\cosh u\right)dx}
{\left[\left(x - \hat b\cosh u\right)^2 + \left(1 - \hat b^2\right)
\sinh^2 u\right]^{3/2}}\;.
\end{equation}
Let $y = x - \hat b\cosh u$; $dy = dx$ (for these purposes $u$ is
fixed):
\begin{eqnarray}
\gamma_1^{\rm NP} &=& \frac{\mu}{M}
\int\frac{\left(\hat b\sinh^2 u + y\cosh u\right)dy}
{\left[y^2 + \left(1 - \hat b^2\right)\sinh^2 u\right]^{3/2}}
\nonumber\\
&=& \frac{\mu}{M}\left[
\int\frac{y\cosh u\,dy}
{\left[y^2 + \left(1 - \hat b^2\right)\sinh^2 u\right]^{3/2}}
+ \int\frac{\hat b\sinh^2 u\,dy}
{\left[y^2 + \left(1 - \hat b^2\right)\sinh^2 u\right]^{3/2}}
\right]\;.
\end{eqnarray}
The first of these integrals is:
\begin{eqnarray}
\int\frac{y\cosh u\,dy}
{\left[y^2 + \left(1 - \hat b^2\right)\sinh^2 u\right]^{3/2}}
&=& -\frac{\cosh u}{\left[y^2 + (1 - \hat b^2)\sinh^2 u\right]^{1/2}}
\nonumber\\
&=& -\frac{\cosh u}{\left[\sinh^2 u\sin^2 v + \left(\hat b - \cosh
u\cos v\right)\right]^{1/2}}\;.
\end{eqnarray}
The second yields
\begin{eqnarray}
\int\frac{\hat b\sinh^2 u\,dy}
{\left[y^2 + \left(1 - \hat b^2\right)\sinh^2 u\right]^{3/2}}
&=& \frac{y\hat b\sinh^2 u}
{\left[\left(1 - \hat b^2\right)\sinh^2 u\right]
\left[y^2 + \left(1 - \hat b^2\right)\sinh^2 u\right]^{1/2}}
\nonumber\\
&=& \frac{\hat b}{1 - \hat b^2}
\frac{\cos v - \hat b\cosh u}
{\left[\sinh^2 u \sin^2 v + \left(\hat b - \cosh u\cos
v\right)^2\right]^{1/2}}\;.
\end{eqnarray}
Restoring factors of $M$, the complete integral is
\begin{equation}
\gamma_1^{\rm NP} = \frac{\mu M\left(b \cos v - M\cosh u\right)}
{\left(M^2 - b^2\right) \left[M^2\sinh^2 u \sin^2 v + \left(b - M\cosh
u\cos v\right)^2\right]^{1/2}}\;.
\label{eq:gammaNP_noconst}
\end{equation}

In these integrals, we have neglected integration constants that must
now be determined.  We do so by requiring that $\gamma^{\rm NP}$ go to
zero at large radius, i.e.\ as $u\to\infty$.  Our current form of the
solution has
\begin{equation}
\gamma_1^{\rm NP}(u\to\infty)
= -\frac{\mu M}{M^2 - b^2}\;.
\label{eq:gammaNP_largeu}
\end{equation}
This is the value that must be subtracted to guarantee that
$\gamma_1^{\rm NP}$ is well behaved at large radius.  We thus finally
obtain
\begin{equation}
\gamma_1^{\rm NP} =
\left(\frac{\mu M}{M^2 - b^2}\right)
\frac{b\cos v - M\cosh u}
{\left[M^2\sinh^2u\sin^2v + \left(b - M\cosh u\cos v\right)^2
\right]^{1/2}} + \frac{\mu M}{M^2 - b^2}\;
\label{eq:gam1_NP_full}.
\end{equation}
To account for the south pole perturbation, we just add the
perturbation with $b\to -b$.  [Note that we do {\it not} add the south
perturbation with an overall minus sign, as is done in Ref.\
{\cite{spr88}}.  That sign choice only holds for $u \ll 1$; see Ref.\
{\cite{spr88}}, discussion following Eq.\ (3.14).  Note that this has
no impact on our ability to study strong field orbits --- the region
$u \ll 1$ corresponds to a region just barely outside the event
horizon, well inside of the last stable orbit.]  By direct
substitution, one can verify that this solution for $\gamma_1$
satisfies Eq.\ (\ref{eq:dgam1du}) as well.

The perturbed metric is now fully specified.  For our purposes, it is
very convenient to use Eq.\ (\ref{eq:schw_coords}) to convert to
Schwarzschild coordinates.  Doing so and defining the function
\begin{equation}
D(r, \theta, b) = \left(r^2 - 2 M r + b^2 + M^2\cos^2\theta -
2 b r\cos\theta + 2 b M\cos\theta\right)^{1/2}\;,
\label{eq:big_D}
\end{equation}
the complete perturbations $\psi_1$ and $\gamma_1$ are
\begin{equation}
\psi_1 = -\frac{\mu}{2 D(r,\theta,b)} - \frac{\mu}{2 D(r,\theta,-b)}\;,
\label{eq:full_psi1_points}
\end{equation}
\begin{equation}
\gamma_1 = \frac{\mu M}{M^2 - b^2}\left[
\left(\frac{M + b\cos\theta - r}{D(r,\theta,b)}\right) +
\left(\frac{M - b\cos\theta - r}{D(r,\theta,-b)}\right)\right]
+ \frac{2\mu M}{M^2 - b^2}\;.
\label{eq:full_gam1_points}
\end{equation}
These expressions hold for all $r > 2M$.  Note the Schwarzschild
coordinate locations of the point masses are $r = b + M$, $\theta =
0,\pi$.  We must choose $b > M$ in order for the perturbing masses to
be in the hole's exterior.

Combining Eqs.\ (\ref{eq:en_of_p_e}), (\ref{eq:angmom_of_p_e}),
(\ref{eq:diff_periapse}), (\ref{eq:periaps_prec}),
(\ref{eq:full_psi1_points}), and (\ref{eq:full_gam1_points}), we can
now calculate the periapse precession for equatorial orbits and the
two-point-mass perturbation.  Numerical results for the general,
strong-field case will be discussed in Sec.\
{\ref{subsec:both_strong}}; here, we show the weak-field result ($p
\gg M$, $p \gg b$).  This weak-field phase shift breaks naturally into
3 pieces:
\begin{equation}
\Delta\phi^{\rm points} = \Delta\phi_{\rm Schw}(M + \mu) +
\Delta\phi_{\rm anom}(\mu) + \Delta\phi_{\rm prol}(\mu, b)\;,
\label{eq:points_weak_dphi}
\end{equation}
where
\begin{eqnarray}
\Delta\phi_{\rm Schw}(M + \mu) &=& \frac{6\pi(M + \mu)}{p} +
\frac{3\pi(M^2 + 2M\mu)}{2p^2}\left(18 + \varepsilon^2\right)\;,
\label{eq:dphi_schw}\\
\Delta\phi_{\rm anom}(\mu) &=& -\frac{\pi\mu M}{p^2}
\left(1 + 2\varepsilon^2\right)\;,
\label{eq:dphi_anom}\\
\Delta\phi_{\rm prol}(\mu,b) &=& -\frac{3\pi\mu b^2}{Mp^2}\;.
\label{eq:dphi_prolate}
\end{eqnarray}
These three pieces each have a simple, physical explanation.  The
first, $\Delta\phi_{\rm Schw}$ is just the periapsis precession
expected for a particle orbiting a Schwarzschild black hole with total
mass $M + \mu$ (to leading order in $\mu$, and to second order in
$1/p$).  The second, $\Delta\phi_{\rm anom}$, is an ``anomalous''
contribution to the precession arising from the fact that our
perturbation is non-spherical and has non-zero total mass.  As we show
later, this term can be eliminated by appropriately designing the
perturbing mass distribution.

The third piece, $\Delta\phi_{\rm prol}$, is the leading order
contribution to the precession that arises from a prolate, quadrupolar
distortion to the black hole.  This term agrees exactly with a
Newtonian calculation of periapse precession in the presence of a
prolate quadrupolar distortion (cf.\ Appendix \ref{sec:newton}).

\section{Metric Perturbations II: An equatorial ring}
\label{sec:ring}

We next examine perturbations due to a circular, equatorial ring of
mass $\mu$ about a black hole.  Our procedure is essentially the same
as that used in the two-point-mass example, though the form of the
perturbing potential makes the calculation a bit more complicated.  We
choose for $\psi_1$ the potential of a ring with radius $\rho = b$ and
total mass at infinity $\mu$:
\begin{equation}
\psi_1^{\rm ring} = -\frac{\mu}{2\pi}\int_0^{2\pi}
\frac{d\xi}{\left[\rho^2 + z^2 + b^2 - 2 b\rho\cos\xi\right]^{1/2}}\;.
\label{eq:psi1_ring}
\end{equation}
This potential can be re-expressed using the complete elliptic
integral of the 1st kind; it turns out to be more convenient to leave
it as written here.

Inserting this into Eq.\ (\ref{eq:dgam1dv_fin}) and integrating yields
\begin{equation}
\gamma_1^{\rm ring} = \int_0^{2\pi}d\xi\left[{\cal C}(\xi) -
\frac{f(u,v,\xi)}{g(u,v,\xi)\ h(u,v,\xi)}\right]\;,
\label{eq:gam1_ring}
\end{equation}
where
\begin{eqnarray}
f(u,v,\xi) &=& \mu\cosh u \left(b^2 + M^2\cosh^2 u - Mb\sinh u\sin
v\cos\xi\right)\;,\\
g(u,v,\xi) &=& \pi\left[\left(M^2 + b^2\cos^2\xi\right)\cosh^2u + b^2
\sin^2\xi\right]\;,\\
h(u,v,\xi) &=& \left(M^2\sinh^2 u\sin^2 v + M^2\cosh^2 u\cos^2v + b^2 -
2 M b \sinh u\sin v\cos\xi\right)^{1/2}\;.
\end{eqnarray}
The integration ``constant'' ${\cal C}(\xi)$ is chosen, as in the
two-point-mass case, to make $\gamma_1$ vanish at large distances.  It
has the value
\begin{equation}
{\cal C}(\xi) = \frac{\mu M}{\pi\left(M^2 + b^2\cos^2\xi\right)}\;.
\end{equation}
To evaluate the periapsis precession, it is useful to have these
results in Schwarzschild coordinates.  Using Eq.\
(\ref{eq:schw_coords}), $f$, $g$, and $h$ become
\begin{eqnarray}
f(r,\theta,\xi) &=& \frac{\mu}{M}(r - M)\left[b^2 + (r - M)^2 - b
\sqrt{r^2 - 2Mr}\sin\theta\cos\xi\right]\;,\\
g(r,\theta,\xi) &=& \pi\left[\left(1 + \frac{b^2}{M^2}\cos^2\xi\right)
(r - M)^2 + b^2\sin^2\xi\right]\;,\\
h(r,\theta,\xi) &=& \left[(r^2 - 2 M r)\sin^2\theta + (r -
M)^2\cos^2\theta + b^2 - 2 b \sqrt{r^2 -
2Mr}\sin\theta\cos\xi\right]^{1/2}\;.
\end{eqnarray}
The ring's radius in Schwarzschild coordinates is $r = M + \sqrt{M^2 +
b^2}$; any choice $b > 0$ will produce a ring in the hole's exterior.

Since the integration over $\xi$ commutes with other operations
(notably expanding for large $p$) it is straightforward to combine
Eqs.\ (\ref{eq:en_of_p_e}), (\ref{eq:angmom_of_p_e}),
(\ref{eq:diff_periapse}), (\ref{eq:periaps_prec}),
(\ref{eq:psi1_ring}), and (\ref{eq:gam1_ring}) to compute the
periapsis precession in this spacetime.  We do so numerically in Sec.\
{\ref{subsec:both_strong}}; for now, we evaluate the weak-field
precession ($p \gg M$, $p \gg b$):
\begin{equation}
\Delta\phi^{\rm ring} = \Delta\phi_{\rm Schw}(M + \mu) +
\Delta\phi_{\rm anom}(\mu) + \Delta\phi_{\rm obl}(\mu, b)\;.
\label{eq:ring_weak_dphi}
\end{equation}
The terms $\Delta\phi_{\rm Schw}(M + \mu)$ and $\Delta\phi_{\rm
anom}(\mu)$ are defined in Eqs.\ (\ref{eq:dphi_schw}) and
(\ref{eq:dphi_anom}); they are identical to the corresponding terms
for the point mass perturbations.  The term
\begin{equation}
\Delta\phi_{\rm obl}(\mu,b) = +\frac{3\pi\mu b^2}{2Mp^2}
\label{eq:dphi_oblate}
\end{equation}
gives the contribution to the precession due to an {\it oblate}
quadrupolar distortion to the black hole.  As in the prolate case,
this term agrees exactly with a Newtonian calculation of periapse
precession presented in Appendix \ref{sec:newton}.

\section{Metric Perturbations III: Pure quadrupole perturbation}
\label{sec:both}

\subsection{Weak-field analysis}
\label{subsec:both_weak}

Each of our perturbations changed not only the quadrupole ($l = 2$)
structure of the spacetime, but the spacetime's total mass ($l = 0$)
as well.  To construct a perturbation that is purely quadrupolar (at
least in the weak field), we take advantage of the linearity of the
perturbed Einstein equations to superpose a ring of mass $\pm\mu$ with
a pair of points of mass $\mp\mu/2$ each.  A negative mass
perturbation may seem strikingly unphysical, but as discussed in Sec.\
{\ref{sec:basics}}, our goal is not to build perturbations that are
likely to exist in nature.  (Given the naked singularity
interpretation, it is arguable that a negative mass perturbation is no
less physical than one of positive mass.)

From now on, we will focus on this zero-mass perturbation.  The weak
field result periapse precession for this case is
\begin{equation}
\Delta\phi^{\rm both} = \Delta\phi_{\rm Schw}(M) + \Delta\phi_{\rm
quad}(\mu,b)\;,
\end{equation}
where the quadrupole contribution is
\begin{equation}
\Delta\phi_{\rm quad}(\mu,b) = \pm\frac{9\pi\mu b^2}{2 M p^2}\;.
\label{eq:dphi_quad}
\end{equation}
Notice that the ``anomalous'' contribution to the precession does not
appear in this case.  Making the perturbation have zero mass
apparently suffices to eliminate that term.  What remains is a
periapsis precession term corresponding to a {\it purely quadrupolar}
deformation of the spacetime.  We have succeeded in constructing a
spacetime that is {\it almost}\ that of a black hole, that is good
deep into the strong field, but has one multipole moment --- $M_2$ ---
with the ``wrong'' value.

\subsection{Strong-field analysis}
\label{subsec:both_strong}

Having demonstrated that our calculations match up in a sensible way
with Newtonian and relativistic calculations of periapse precession,
we now show numerical calculations for the strong fields of bumpy
black holes.  In all of our figures and discussion, our focus is on
the precession induced {\it relative to}\ periapsis precession for a
normal black hole.  To help calibrate the effect that the bumpiness
has, we show in Fig.\ {\ref{fig:schw}} the periapsis precession
$\delta\phi$ (the amount of azimuth $\phi$ accumulated in one orbit,
minus $2\pi$) in the strong field of a Schwarzschild black hole.
Notice that $\delta\phi$ becomes quite large in the strong field,
corresponding to almost 2 extra revolutions around the hole.

Figures {\ref{fig:b2.5}} and {\ref{fig:b1.5}} show the shift
$\Delta\phi$ associated with two different choices of bumpiness.  In
both plots, we have put $\mu = 0.01$; since the Einstein equations and
the equations of motion all scale linearly with $\mu$, $\Delta\phi$ is
likewise linear in $\mu$.  Thus, it is a simple matter to rescale to
other masses.  We examine strong-field precession for $\varepsilon =
0.7$ and $\varepsilon = 0.1$ in these plots; we also compare these
results to the weak-field prediction (\ref{eq:dphi_quad}).

Several features are evident.  First, in both cases, the strong-field
results asymptotically approach the weak-field formula for large $p$
--- an important sanity check.  Interestingly, as $b$ is increased
from $1.5M$ to $2.5M$, the weak-field formula changes from
overestimating the periapsis shift to underestimating it.  This
behavior may be due to deviations of our potential from that of a
perfect quadrupole.  Second, at most orbits, the eccentricity has very
little impact on this shift, at least for the values we examine.  This
is not surprising; the bare Schwarzschild periapsis precession (Fig.\
{\ref{fig:schw}}) is likewise fairly insensitive to the eccentricity.
Finally, even into the strong field, the periapsis shift scales
approximately with $b^2$ (the weak-field prediction).  We can thus
regard $\mu b^2/M^3$ as a measurable, dimensionless ``bumpiness
parameter'' for the black hole.

Most interesting is the robustness with which the quadrupolar bump is
manifested in the periapsis result: the effect is quite pronounced in
the strong field.  Indeed, the weak-field prediction
(\ref{eq:dphi_quad}) underestimates the degree of precession due to the
hole's bumpiness by a factor $\sim3$ over much of the strong field
($\sim 10$ as we approach the last stable orbits).  Over much of the
strong field, the effect is large enough that, by observing over
multiple orbital cycles, we should be able to set very interesting
limits on the bumpiness of black hole candidates.  For example, at $p
\sim 20 M$, the periapsis shift per orbit is roughly
\begin{equation}
\Delta\phi \simeq 10^{-3}\left(\frac{\mu b^2/M^3}{0.02}\right)\;.
\end{equation}
A bumpiness $\mu b^2/M^3\simeq 0.02$ should thus be easily measurable
after tracking roughly $1000$ orbits (corresponding to when the bump
shifts the accumulated phase by about 1 radian).  Conversely, tracking
the phase for $N$ orbital cycles at this value of $p$ should make it
possible to constrain the bumpiness to be
\begin{equation}
\frac{\mu b^2}{M^3} \lesssim \frac{20}{N}\;.
\end{equation}
Better limits can be obtained for orbits at smaller $p$.

Obviously, more detailed work is needed to carefully examine how well
black hole bumpiness can be measured in a variety of scenarios (akin
to Ryan's analysis of multipole measurability by LISA, Ref.\
{\cite{r_measure}).  But, these results suggest that measurements
which coherently follow orbital phases --- such as gravitational-wave
measurements and x-ray timing --- should be able to place {\it
stringent} constraints on the bumpiness of black hole candidates.
Bumpy black holes should thus be very useful tools in designing a
formalism to map the strong field structure of black hole candidates
in nature.

\section{Conclusion}
\label{sec:concl}

In this paper, we have laid the foundations for a null experiment to
test whether a massive compact object is a black hole.  The bumpy
black hole spacetimes we construct differ only slightly from normal
black hole spacetimes; and, the difference is controlled by a simple
adjustable parameter --- the hole's ``bumpiness''.  It should be
possible to compare the properties of black hole candidates in nature
with these bumpy black hole spacetimes.  If these objects are in fact
the black holes of general relativity, measurements will show that the
natural spacetimes have a bumpiness of zero.

Quite a bit more work is needed in order to make the bumpy black hole
concept useful in practice for astrophysical measurements:

\begin{itemize}

\item Foremost is the need to generalize this analysis to bumpy Kerr
black holes --- zero angular momentum is a highly unrealistic
idealization.  In Appendix {\ref{sec:kerr}}, we show how, by choosing
$\psi$, $\gamma$, and using an appropriate coordinate transformation,
the Weyl metric (\ref{eq:weyl_cyl}) encompasses Kerr black holes (in
Boyer-Lindquist coordinates).  There should then be no severe
difficulty perturbing this metric to build bumpy Kerr black holes,
though the details are likely to be complicated.

\item Probably next in importance is generalizing the orbits which we
analyze to include inclination with respect to the hole's equatorial
plane.  Besides the astrophysical motivation --- we do not often
expect orbits to be confined to a special plane --- the inclusion of
an extra degree of orbital freedom offers opportunity.  Motions out of
the plane are characterized by oscillations with a frequency
$\Omega_\theta$ which is generically different from the frequencies
$\Omega_\phi$ and $\Omega_r$ discussed in this paper.  These
oscillations thus offer an additional ``handle'' by which deviations
from the black hole spacetime can be characterized.

As already mentioned in Sec.\ {\ref{sec:orbits}}, the equations of
motion for inclined orbits of bumpy black holes do not appear to
separate (as they do for normal Kerr black holes).  However, the fact
that, by definition, bumpy black hole spacetimes are {\it nearly}\ the
spacetimes of black holes suggests that the equations of motion must
{\it nearly} separate.  In other words, the degree to which the
$\theta$ motion couples to the $r$ and $\phi$ motion must be small ---
no doubt controlled by a coupling factor that is of order $\mu/M$.  It
may be possible to take advantage of this smallness to usefully
describe inclined and eccentric bumpy black hole orbits.

\item We have focused on perturbing mass distributions that produce,
in the weak field, a purely quadrupolar spacetime distortion.  We
chose to focus on this case because an incorrect quadrupole moment is
sufficient to falsify the black hole hypothesis.  There is no reason
why we could not go beyond this: by using rings out of the hole's
equatorial plane, one could imagine building essentially any
multipolar distribution whatsoever.  Indeed, the fact that the
equation for the metric function $\psi$ in Weyl coordinates [Eq.\
(\ref{eq:d2psidrho2})] is simply $\nabla^2\psi = 0$ tells us that it
is simple in principle to specify perturbations whose weak-field
multipolar structure is completely arbitrary: the perturbation
\begin{equation}
\psi_1 = \sum \frac{{\cal B}_l Y_{l0}}{(\rho^2 + z^2)^{(l+1)/2}}
\end{equation}
will work perfectly.  The parameter ${\cal B}_l$ is a generalized
$l$-polar ``bumpiness''.  With this ansatz to define our deviations,
we can build a bumpy black holes with almost arbitrarily shaped bumps.
This will make it possible to strongly constrain the properties of
black holes in nature.

As this paper was being completed, an analysis appeared on the gr-qc
archive by Ashtekar et al.\ {\cite{ih_multipoles}} of the multipole
moments of isolated horizons {\cite{ih_prl}}.  Although we have not
investigated this in any depth, it may be beneficial to pursue a
connection between the bumpiness of a black hole and the multipole
moments expressed in the language of Ref.\ {\cite{ih_multipoles}}.

\end{itemize}

With these generalizations in hand, it should be possible to begin
examining detailed mechanisms by which orbital frequencies can be
imprinted on astrophysical observables.  For example, one can imagine
analyzing accretion disk models to see how a spacetime's bumpiness is
imprinted on quasi-periodic oscillations in a source's x-ray spectrum
{\cite{psaltis,sb04,r04}}.  Another example is in gravitational-wave
science.  For these ideas to be useful for testing the nature of black
hole candidates, we will need to model the gravitational-wave emission
and inspiral of compact bodies captured by bumpy black holes.  This
problem may not be much more difficult than the corresponding problem
for normal Kerr black holes --- because the wave amplitude is itself
perturbatively small, the inspiral and wave generation should decouple
(at least to first order) from the spacetime's bumps.  We hope to
address at least some of these issues in future work.

\acknowledgments

We thank Deepto Chakrabarty and \'Eanna Flanagan for useful
discussions.  The package {\sc mathematica} was used to aid some of
the calculations, particularly the weak-field expansions.  This work
was supported by NASA Grant NAG5-12906 and NSF Grant PHY-0244424.

\appendix

\section{Periapse Precession in Newtonian Theory}
\label{sec:newton}

An elliptical orbit precesses even in Newtonian gravity if the orbit
is about a body with a quadrupole moment.  Here, we calculate this
precession, presenting results in a form useful for making contact
with this paper's relativistic results.

A body with a mass and a quadrupole moment has a Newtonian
gravitational potential
\begin{equation}
\Phi = -\frac{M}{r} -\frac{3}{2}\frac{{\cal I}_{ab}n^a n^b}{r^3}\;.
\label{eq:potential}
\end{equation}
For further discussion, see Thorne's voluminous treatise on multipole
moments in general relativity {\cite{t80}}.  The tensor ${\cal
I}_{ab}$ is the symmetric, trace-free quadrupole moment of the
gravitating source:
\begin{equation}
{\cal I}_{ab} = \int d^3r\,\rho\,\left(x_a x_b -
\frac{1}{3}\delta_{ab} r^2\right)\;.
\label{eq:STFquad}
\end{equation}
The vector $n^a$ is a direction cosine.  For notational simplicity, we
put ${\cal I}_{ab}n^a n^b \equiv Q$.

A body in an equatorial orbit around this object obeys a simple
equation of motion:
\begin{eqnarray}
\left(\frac{dr}{dt}\right)^2 &=& E - \Phi - \frac{L^2}{2 r^2}
\nonumber\\
&\equiv& \frac{R_{\rm newt}(r)}{r^3}\,,
\label{eq:eom}
\end{eqnarray}
where $L$ is the component of orbital angular momentum perpendicular
to the equatorial plane. We are interested in eccentric orbits, so we
reparameterize in the usual way:
\begin{equation}
r = \frac{p}{1 + \varepsilon\cos\chi}\;,
\end{equation}
which is equivalent to
\begin{equation}
u = v(1 + e\cos\chi)
\label{eq:u_eccentric}
\end{equation}
with $u = 1/r$, $v = 1/p$. The turning points of an eccentric orbit
are at apoapsis, $u_a = v(1 - \varepsilon)$, and periapsis, $u_p = v(1
+ \varepsilon)$.  Solving the equation of motion at both turning
points --- $R_{\rm newt}(u_a) = 0$, $R_{\rm newt}(u_p) = 0$ --- yields
a solution for the energy $E$ and angular momentum $L$ as functions of
$v$ and $\varepsilon$:
\begin{eqnarray}
E &=& \frac{1}{2}\left(\varepsilon^2 - 1\right) M v +
\frac{3}{4}\left(\varepsilon^2 - 1\right)^2 Qv^3\;,
\label{eq:energy}\\
L^2 &=& \frac{M}{v} + \frac{3}{2}\left(3 + \varepsilon^2\right) Q v\;.
\label{eq:angmom}
\end{eqnarray}
An orbit corresponds to moving through the range $\chi = 0$ to $\chi =
2\pi$.  We want to find the amount of azimuthal angle $\phi$ that
accumulates over that orbit; that number (minus $2\pi$) is the
accumulated periapsis precession.  First, use the above results for
$E$ and $L$ in the radial equation of motion:
\begin{equation}
\left(\frac{dr}{dt}\right)^2 = \frac{1}{2}\varepsilon^2
v\sin^2\chi\left[2 M + 3\left(\varepsilon^2 - 3\right)Q v^2 - 6
\varepsilon Q v^2\cos\chi\right]
\label{eq:drdt}
\end{equation}
Next, using $r = 1/v(1 + \varepsilon\cos\chi)$ we have
\begin{equation}
\frac{dr}{dt} = r^2\varepsilon v\sin\chi\frac{d\chi}{dt}\;.
\label{eq:drdxi}
\end{equation}
We obtain
\begin{equation}
\left(\frac{d\chi}{dt}\right)^2 = \frac{2 M + 3\left(\varepsilon^2 -
3\right)Q v - 6\varepsilon Q v\cos\chi}{2 v r^4}\;.
\label{eq:dxidt}
\end{equation}
We leave the $r^4$ in the denominator to cancel another factor that
will appear shortly.

To connect this to the azimuthal angle $\phi$, we use the fact that $L
= r^2\,d\phi/dt$.  Combining this with (\ref{eq:dxidt}), we obtain an
equation for the differential periapsis advance:
\begin{eqnarray}
\left(\frac{d\phi}{d\chi}\right)^2
&=& \left(\frac{d\phi}{dt}\right)^2 \left(\frac{d\chi}{dt}\right)^{-2}
\nonumber\\
&=& \left(\frac{L}{r^2}\right)^2 \left[\frac{2 v r^4}{2 M +
3(\varepsilon^2 - 3)Q v - 6\varepsilon Q v\cos\chi}\right]
\nonumber\\
&=& \frac{2 M + 3 (3 + \varepsilon^2) Q v^2}{2 M + 3(\varepsilon^2 -
3) Q v^2 - 6 \varepsilon Q v^2\cos\chi}
\end{eqnarray}

We now take the square root.  We assume that $Q v^2 \equiv Q/p^2 \ll
M$ and expand in $v$.  This amounts to a weak field expansion.  We
find, to leading order in $Q v^2/M$,
\begin{equation}
\frac{d\phi}{d\chi} = 1 + \frac{3 Q v^2(3 + \varepsilon\cos\chi)}{2M}
\label{eq:advance}
\end{equation}
Integrating over an orbit yields
\begin{equation}
\Delta\phi = \int_0^{2\pi}d\chi\frac{d\phi}{d\chi} - 2\pi = \frac{9\pi
Q v^2}{M}\;.
\label{eq:shift_Q}
\end{equation}

We now compute $Q$ for the cases examined in the text: two point
masses on the symmetry axis, and an equatorial ring.  The direction
cosines are easy to calculate; we are only interested in orbits on the
equatorial plane, so $n^z = 0$.  By axisymmetry, all directions in the
equatorial plane are equivalent, so we choose $n^x = 1$, $n^y =
0$. Then, $Q = {\cal I}_{xx}$.

For a point mass $\mu/2$ at the north pole,
\begin{equation}
{\cal I}^{\rm NP}_{ab} = \frac{\mu b^2}{6}{\rm diag}(-1,-1,2)\;;
\end{equation}
for a point mass at the south pole, we find the same result.  The pair
of point masses thus has $Q = -\mu b^2/3$.  Using this $Q$ in Eq.\
(\ref{eq:shift_Q}) we find
\begin{equation}
\Delta\phi^{\rm points} = -\frac{3\pi\mu b^2}{p^2M}\;.
\end{equation}

Repeating this exercise for the ring, we find
\begin{equation}
{\cal I}^{\rm ring}_{ab} = \frac{\mu b^2}{6}{\rm diag}(1,1,-2)\;,
\end{equation}
so $Q = \mu b^2/6$, and
\begin{equation}
\Delta\phi^{\rm ring} = +\frac{3\pi\mu b^2}{2p^2M}\;.
\end{equation}
Notice that $\Delta\phi^{\rm points} = -2\Delta\phi^{\rm ring}$.  This
follows from the spherical harmonics that describe the ring
[$Y_{20}(\pi/2)$] and the point masses [$Y_{20}(0) =
-2Y_{20}(\pi/2)$].

\section{The Kerr metric from the Weyl metric}
\label{sec:kerr}

In this appendix, we lay out the coordinate transformations and the
choices of $\psi$ and $\gamma$ needed to go from the Weyl metric to
the Kerr metric in Boyer-Lindquist coordinates.  Although only the
uncharged version is astrophysically relevant, we show the results for
general charge $Q$.  This calculation shares much with the
Newman-Janis algorithm {\cite{nj65,netal65,ds2000}}.

We begin with the Weyl metric in prolate spheroidal coordinates, Eq.\
(\ref{eq:weyl_prolate}).  We then choose
\begin{equation}
M\cosh u = r - M + i \sqrt{a^2 + Q^2}\;.
\end{equation}
Note that $u$ must be complex; the black hole parameters $M$, $a$, and
$Q$, as well as the radial coordinate $r$ are of course real.  As we
discuss further below, this implies a severe constraint on the values
$u$ may take in the complex plane.

It follows that
\begin{eqnarray}
dr &=& M\sinh u\ du\;,
\nonumber\\
M^2 \cosh^2 u &\equiv& M^2 |\cosh u|^2
\nonumber\\
&=& (r - M)^2 + a^2 + Q^2\;,
\nonumber\\
M^2 \sinh^2 u &=& r^2 - 2 M r + a^2 + Q^2\;.
\end{eqnarray}
We also choose $v = \theta$.  For notational convenience, define
\begin{eqnarray}
\Delta &=& r^2 - 2 M r + a^2 + Q^2\;,
\nonumber\\
\Sigma &=& r^2 + a^2\cos^2\theta\;.
\end{eqnarray}
Substituting into the metric (\ref{eq:weyl_prolate}), we find
\begin{eqnarray}
ds^2 &=& -e^{2\psi}\,dt^2
+ e^{2\gamma - 2\psi}
\left(\frac{\Delta + M^2\sin^2\theta}{\Delta}\right)dr^2
+ e^{2\gamma - 2\psi}\left(\Delta + M^2\sin^2\theta\right)d\theta^2
+ e^{-2\psi}\Delta \sin^2\theta\,d\phi^2\;.
\end{eqnarray}

Next, choose
\begin{eqnarray}
e^{2\psi} &=& \frac{\Delta}{\Sigma}\;,
\nonumber\\
e^{2\gamma} &=& \frac{\Delta}{\Delta + M^2\sin^2\theta}\;,
\end{eqnarray}
and put
\begin{eqnarray}
dt &=& dt' - a\sin^2\theta\,d\phi',
\nonumber\\
d\phi &=& \frac{(r^2 + a^2)}{\Sigma}\,d\phi' - \frac{a}{\Sigma}\,dt'\;.
\end{eqnarray}
Dropping the primes on $t$ and $\phi$, we see that the Weyl metric
reduces to the Kerr-Newman metric in Boyer-Lindquist coordinates:
\begin{equation}
ds^2 = -\frac{\Delta}{\Sigma}\left(dt -
a\sin^2\theta\,d\phi\right)^2
+ \frac{\sin^2\theta}{\Sigma} \left[\left(r^2 + a^2\right)d\phi -
a\,dt\right]^2
+ \frac{\Sigma}{\Delta}\,dr^2 + \Sigma\,d\theta^2\;.
\end{equation}
Compare, for example Ref.\ {\cite{mtw}}, Eq.\ (33.2) and Ref.\
{\cite{wald}}, Eq.\ (12.3.1).

In prolate spheroidal coordinates, the functions $\psi$ and $\gamma$
are given by
\begin{eqnarray}
e^{2\psi} &=& \frac{M^2\cosh^2u}{r^2(u) + a^2\cos^2 v}\;,
\\
e^{2\gamma} &=& \frac{M^2\cosh^2u}{M^2\cosh^2u + M^2\sin^2 v}\;,
\end{eqnarray}
where
\begin{equation}
r(u) = M\left(\cosh u - 1\right) - i\sqrt{a^2 + Q^2}\;.
\end{equation}
The requirement that $r(u)$ be real implies a severe constraint
between the real and imaginary parts of $u$: we must have ${\rm
Im}(M\cosh u) = +i\sqrt{a^2 + Q^2}$.  Writing $u = u_r + i u_i$,
we find
\begin{equation}
M\sin\left(u_i\right)\sinh\left(u_r\right) = \sqrt{a^2 + Q^2}\;.
\label{eq:u_traj}
\end{equation}
In other words, $u$ cannot take just any value in the complex plane,
but must be confined to the trajectory defined by Eq.\
(\ref{eq:u_traj}).

Using these results, it should straightforward to generalize our
calculations to describe bumpy Kerr black holes.

\begin{figure}
\epsfig{file=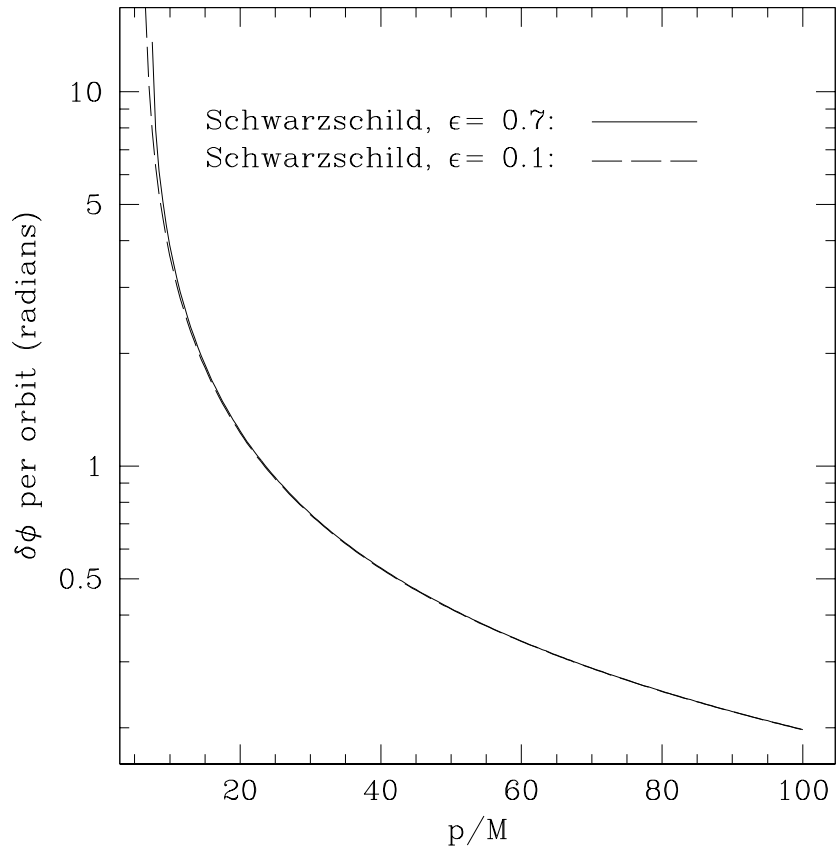, width = 12cm}
\caption{Periapsis precession in the strong field of a Schwarzschild
black hole: the ``extra'' azimuth accumulated over a single orbit with
parameters $(p,\varepsilon)$.  The solid curve is for eccentricity
$\varepsilon = 0.7$; the dashed one is for $\varepsilon = 0.1$.  As we
go to the weak field, $\delta\phi$ approaches zero --- orbits approach
closed ellipses.  The degree of precession is large in the strong
field: the limiting value $\sim 12$ radians corresponds to almost 2
extra revolutions around the symmetry axis.}
\label{fig:schw}
\end{figure}

\begin{figure}
\epsfig{file=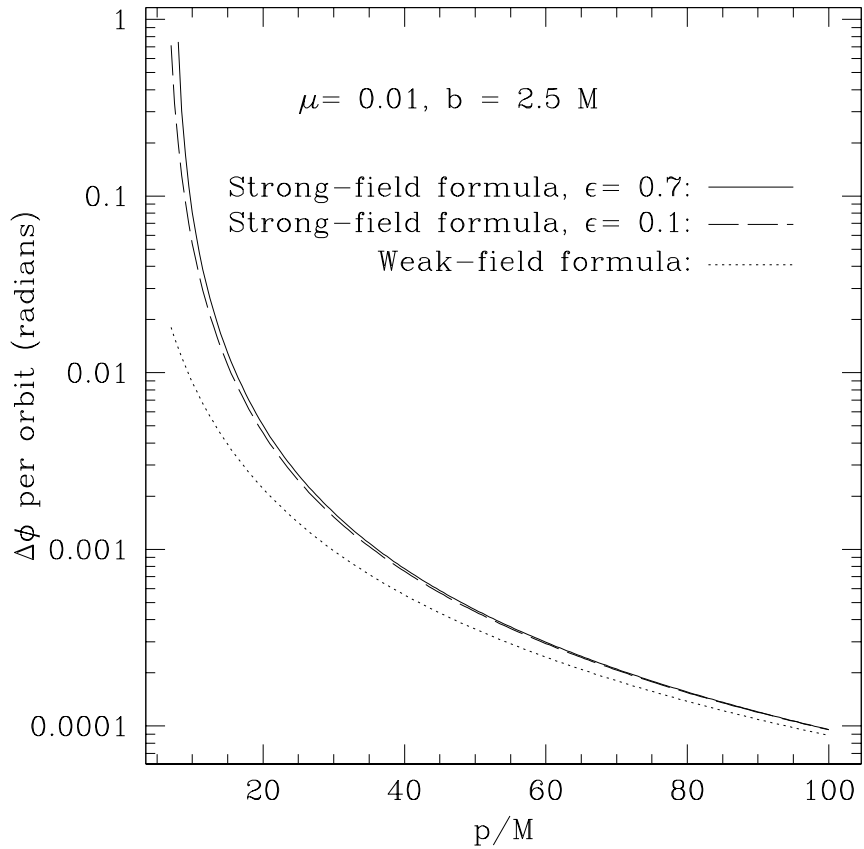, width = 12cm}
\caption{Periapsis precession in the strong field of a bumpy black
hole: the ``extra'' azimuth accumulated due to the hole's bumpiness as
a function of $(p,\varepsilon)$.  We have put $\mu = 0.01$ and $b =
2.5$.  By the linearity of all relevant equations in mass, the degree
of precession scales proportional to $\mu$.  The solid curve is for
$\varepsilon = 0.7$; the dashed one is for $\varepsilon = 0.1$.  The
dotted curve shows the weak-field prediction.  Both strong-field
results asymptotically approach the weak-field prediction for large
$p$.}
\label{fig:b2.5}
\end{figure}

\begin{figure}
\epsfig{file=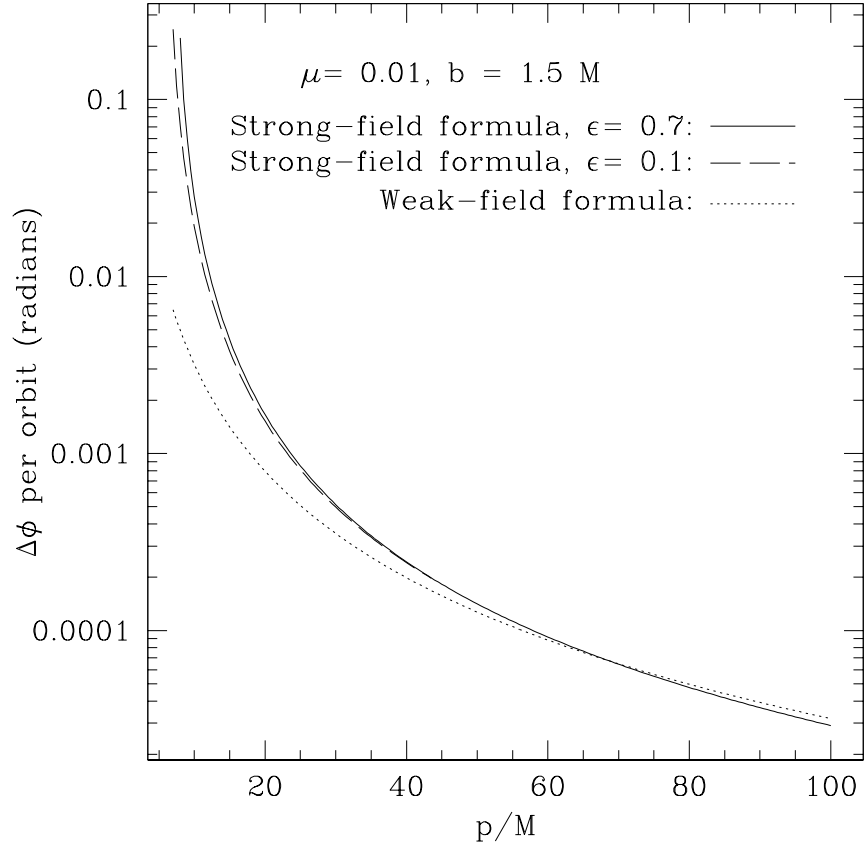, width = 12cm}
\caption{Periapsis precession in the strong field of a bumpy black
hole: the ``extra'' azimuth accumulated due to the hole's bumpiness as
a function of $(p,\varepsilon)$.  This plot is identical to Fig.\
{\ref{fig:b2.5}} except that we have put $b = 1.5M$.}
\label{fig:b1.5}
\end{figure}

\end{document}